\documentclass[12pt]{revtex4}

\usepackage{amsfonts}
\usepackage{amsmath}
\usepackage{amssymb}
\usepackage{graphicx}
\usepackage[usenames]{color}

\begin{document}

\title{Extended Thermodynamic Relation and Fluctuation Theorem in Stochastic Dynamics with Time Reversed Process}
\author{T. Koide}
\affiliation{FIAS, Johann Wolfgang Goethe-University\"at, Ruth-Moufang Str. 1,
60438, Frankfurt am Main, Germany}
\author{M. Mine}
\affiliation{
Waseda University Honjo Senior High School,
1136 Nishitomida, Honjo City, Saitama 367-0035, Japan
}
\author{M. Okumura}
\affiliation{
CCSE, Japan Atomic Energy Agency, 6-9-3 Higashi-Ueno, Taito-ku, Tokyo 110-0015, Japan 
}
\affiliation{
CREST (JST), 4-1-8 Honcho, Kawaguchi-shi, Saitama 332-0012, Japan 
}
\author{Y. Yamanaka}
\affiliation{Department of Electronic and Photonic Systems,
Waseda University, Tokyo 169-8555, Japan}

\begin{abstract}
We consider a stochastic model described by two stochastic differential equations of motion; 
one is for the stochastic evolution forward in time and the other for backward in time.
We further introduce averaged quantities for the two processes and 
construct the extended thermodynamic relation following the strategy of Sekimoto \cite{seki}.
By using this relation, we derive the fluctuation theorems such as 
the Seifert relation, the Jarzynski relation and the Komatsu-Nakagawa non-equilibrium steady state with respect to the  introduced averaged quantities.
\end{abstract}

\maketitle

\section{introduction}

In discussing non-equilibrium processes, it is sometimes important to know 
time evolutions not only forward in time, but also backward in time.
The typical example is the fluctuation theorems, where several non-trivial exact relations are obtained by 
using the different behaviors of the forward and backward processes.

There are mainly two different approaches to discuss the fluctuation theorems; 
one is the discussion based on deterministic dynamics such as the Liouville equation, 
and the other is the stochastic approach using the Langevin equation.

In classical deterministic dynamics, both of the forward and backward evolutions in time are determined by 
the same equation of motion by replacing $t \rightarrow -t$.
However, we have to remember that the time-reversed processes in stochastic dynamics is not trivial 
because the simple manipulation, $t \rightarrow -t$, does not work. 
Let us, for example, consider the free Brownian motion described by 
\begin{eqnarray}
d {\bf X}(t) = \sqrt{2\beta^{-1}} d{\bf W}(t), \nonumber
\end{eqnarray}
where $\sqrt{2\beta^{-1}} d{\bf W}(t)$ is a Gaussian white noise.
It is known that this equation describes the diffusion process, where 
the particle is transfered from a region of higher concentration to one of lower concentration.
Thus the inverse process must be aggregation, which cannot be obtained by changing $t \rightarrow -t$ 
in the above equation.
As a matter of fact, as we will see soon later, the time reversed process is described by
\begin{eqnarray}
d {\bf X}(t) =  \frac{{\bf X}(t)}{t} dt + \sqrt{2\beta^{-1}} d {\bf W}^*(t). \nonumber
\end{eqnarray} 
That is, we need one more equation to set up a stochastic model that describes time-reversed processes.

The purpose of the paper is to investigate  
the fluctuation theorems of this stochastic model.
As we have mentioned, the time reversed process is important to discuss the fluctuation theorems.
However, as far as we know, 
this non-triviality of the time-reversed process in stochastic dynamics 
has not been discussed. 
For this purpose, we first generalize the thermodynamic interpretation of 
stochastic dynamics following the strategy proposed by Sekimoto \cite{seki}, and 
derive the extended thermodynamic relation.
Afterwards, we apply our stochastic model to derive the fluctuation theorems, such as 
the Seifert relation \cite{seifert}, the Jarzynski relation \cite{jar} 
and the Komatsu-Nakagawa non-equilibrium steady state \cite{kn}.

This paper is organized as follows.
In Sec. 2, we introduce our stochastic model.
In Sec. 3, we develop the thermodynamic interpretation of the stochastic dynamics following Ref.~\cite{seki}.
The various fluctuation theorems are discussed in Sec. 4.
Section 5 is devoted to the concluding remarks.

\section{Time Reversed Process in Stochastic Dynamics}

We consider a stochastic process ${\bf X}(t)$ described by the following 
stochastic difference equation, 
\begin{eqnarray}
\gamma d {\bf X}(t) &=& {\bf u}({\bf X}(t),a,t)dt + \sqrt{2\beta^{-1}} d{\bf W}(t)~~~~~~~(dt > 0), \label{langevin1}
\end{eqnarray}
where ${\bf u}$ is a force term, $d{\bf X}(t) \equiv {\bf X}(t+dt) - {\bf X}(t)$, and the noise term satisfies the 
following properties,
\begin{eqnarray}
\langle d{\bf W}(t) \rangle &=& 0,~~~~\langle d{\bf W}_i (t) d{\bf W}_j (t) \rangle = \delta_{ij}|dt|. 
\end{eqnarray}
Here $a$ represents an external control parameter such as an external field.
The parameters $\gamma$ and $\beta$ characterize the magnitude of the relaxation and the noise, respectively.
We consider that $\beta$ is constant.
In the following calculation, the parameter $\gamma$ does not play any important role,  
and, for the sake of simplicity, we use the special unit of $\gamma = 1$.

To discuss the fluctuation theorems, we need to define 
the time reversed process of this stochastic time evolution.
In the classical deterministic dynamics where the noise term $d{\bf W}(t)$ vanishes, 
the time-reversed process is described by replacing $t$ with $-t$ in Eq. (\ref{langevin1}).
However, the time reversed process of the stochastic dynamics is not obtained by 
this simple replacement because of the noise term $d{\bf W}(t)$.

For this purpose, we introduce one more stochastic difference equation which describes the stochastic evolution backward in time, 
\begin{eqnarray}
d {\bf X}(t) &=& {\bf u}^* ({\bf X}(t),a,t)dt + \sqrt{2\beta^{-1}} d {\bf W}^*(t)~~~~(dt < 0), \label{langevin2}
\end{eqnarray}
where 
\begin{eqnarray}
\langle d{\bf W}^*(t) \rangle &=& 0,~~~~\langle d{\bf W}^*_i (t) d{\bf W}^*_j (t) \rangle = \delta_{ij}|dt| ,
\end{eqnarray}
and $d{\bf W}$ and $d{\bf W}^*$ do not have any correlation.
Because $dt$ is negative, this equation describes the backward process.

From the stochastic equations (\ref{langevin1}) and (\ref{langevin2}), 
we obtain the two Fokker-Planck equations \cite{gardiner},
\begin{eqnarray}
\partial_t \rho({\bf r},a,t) &=& - \nabla \cdot ({\bf u}({\bf r},a,t)\rho({\bf r},t)) + \beta^{-1} \nabla^2 \rho({\bf r},a,t), 
\label{fp1}\\
\partial_t \rho({\bf r},a,t) &=& - \nabla \cdot ({\bf u}^*({\bf r},a,t)\rho({\bf r},t)) - \beta^{-1} \nabla^2 \rho({\bf r},a,t).
\end{eqnarray}
When Eq. (\ref{langevin2}) describes the time-reversed process of Eq. (\ref{langevin1}), 
the two Fokker-Planck equations should be equivalent, and 
consequently, ${\bf u}$ and ${\bf u}^*$ are not independent anymore.
As a matter of fact, the two equations are rewritten as 
\begin{eqnarray}
\partial_t \rho({\bf r},a,t) &=& \nabla\cdot (\rho ({\bf r},a,t) {\bf v}({\bf r},a,t))= \nabla \cdot {\bf J}({\bf r},a,t) , \\
\nabla\cdot (\rho({\bf r},a,t){\bf b}({\bf r},a,t)) &=& \beta^{-1} \nabla^2 \rho({\bf r},a,t),
\end{eqnarray}
where
\begin{eqnarray}
{\bf v}({\bf r},a,t) &=& \frac{1}{2}\{ {\bf u}({\bf r},a,t) + {\bf u}^* ({\bf r},a,t) \} , \\
{\bf b}({\bf r},a,t) &=& \frac{1}{2}\{ {\bf u}({\bf r},a,t) - {\bf u}^* ({\bf r},a,t) \}. \label{def_b}
\end{eqnarray}
The solution of the second equation is 
\begin{equation}
{\bf b}({\bf r},a,t) = \beta^{-1} \nabla \ln \rho({\bf r},a,t) + \frac{{\bf c}(a,t)}{\rho({\bf r},a,t)} + \frac{\nabla \times {\bf A}({\bf r},a,t)}{\rho({\bf r},a,t)} .\label{bsol}
\end{equation}
Here ${\bf c}(a,t)$ and ${\bf A}({\bf r},a,t)$ are an arbitrary scalar and vector functions, respectively. 
We set ${\bf c}(a,t)=0$ and $\nabla \times \bf{A}({\bf r},a,t) = 0$ for the sake of simplicity.
Then ${\bf u}^*$ is related to ${\bf u}$ by the following relation, 
\begin{equation}
{\bf u}^* ({\bf r},a,t) = {\bf u}({\bf r},a,t)  - 2\beta^{-1} \nabla \ln \rho({\bf r},a,t). \label{cc}
\end{equation}
It should be emphasized that, as we will see soon later, 
this relationship is the origin of the fluctuation theorems.

In short, the forward and backward stochastic equations of motion are given by  
\begin{eqnarray}
d {\bf X}(t) &=& {\bf u}({\bf X}(t),a,t)dt + \sqrt{2\beta^{-1}} d{\bf W}(t)~~~~~~~~~~~~~~~~~~~~~~~~~~~~~~~~~~~~~~(dt > 0), \\
d {\bf X}(t) &=& [{\bf u} ({\bf X}(t),a,t) - 2\beta^{-1} \nabla \ln \rho({\bf X}(t),a,t) ]dt + \sqrt{2\beta^{-1}} d {\bf W}^*(t)~~~~(dt < 0),
\end{eqnarray} 
respectively.

We apply this stochastic model to, for example, the free Brownian motion, where the forward stochastic equation of motion is 
\begin{equation}
d {\bf X}(t) =  \sqrt{2\beta^{-1}} d{\bf W}(t)~~~~~~~(dt > 0). \label{fbm}
\end{equation}
When we assume $\rho({\bf r},0) = \delta ({\bf r})$, 
the solution of the Fokker-Planck equation (\ref{fp1}) is given by \cite{gardiner}
\begin{equation}
\rho ({\bf r},t) = \frac{e^{-{\bf r}^2/(4\beta^{-1}t) }}{\sqrt{4\pi \beta^{-1} t}}. \label{dm_brown}
\end{equation}
On the other hand, the backward stochastic equation of motion is given by 
\begin{equation}
d {\bf X}(t) =  \frac{{\bf X}(t)}{t} dt + \sqrt{2\beta^{-1}} d {\bf W}^*(t)~~~~(dt < 0). 
\end{equation}
The force term $\frac{{\bf X}(t)}{t}$ gives rise to aggregation overcoming diffusion due to the noise $\sqrt{2\beta^{-1}} d {\bf W}^*(t)$.
We checked numerically that the above equation describes the time-reversed process 
of the free Brownian motion (\ref{fbm}).

\section{Extended Thermodynamic Relation}

In order to discuss the fluctuation theorems, 
it is necessary to know the thermodynamic interpretation in our stochastic model.

The thermodynamic interpretation of the stochastic dynamics without time-reversed processes was 
proposed by Sekimoto \cite{seki}.
When the stochastic dynamics is given by Eq. (\ref{langevin1}), Sekimoto proposed the following relation,
\begin{equation}
dU({\bf X}(t),a) = \delta Q({\bf X}(t),a) + \delta { \cal W} ({\bf X}(t),a), \label{thermo_seki}
\end{equation}
where $U$ is the energy of a Brownian particle defined by 
\begin{equation}
{\bf u}({\bf r},a,t) = -\nabla U({\bf r},a,t),
\end{equation}
and 
\begin{eqnarray}
\delta Q({\bf X}(t),a) &=& \nabla U({\bf X}(t),a) \circ d{\bf X}(t), \label{def_dq_seki}\\
\delta {\cal W} ({\bf X}(t),a) &=& \partial_a U({\bf X}(t),a) \circ da .
\end{eqnarray}
Here, the product $\circ$ means the Stratonovich definition of the stochastic product \cite{gardiner}.
The equation (\ref{thermo_seki}) is obtained by the stochastic differentiation along the stochastic trajectory 
$\{ {\bf X}(t) \}$. In this sense, this equation is an identity. 
Sekimoto assigned the thermodynamic interpretation for the r.h.s. of the identity; 
$\delta Q$ is the heat which a Brownian particle receives 
from the environment, and $\delta {\cal W}$ is 
the work which is done to a Brownian particle from the environment.
This thermodynamic relation has been studied from various points of view \cite{seifert,various}.

\begin{figure}[tbp]
\includegraphics[scale=0.25]{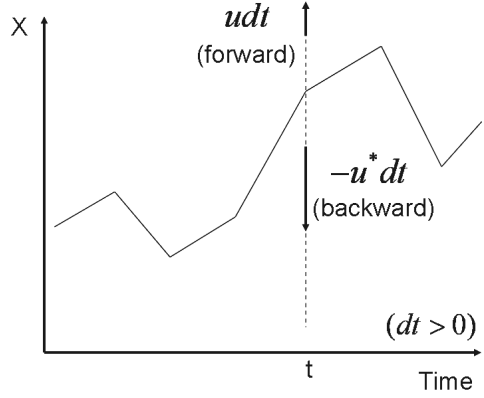} 
\caption{The forces acts on a Brownian particle at $t$. We set $dt > 0$.}
\label{fig1}
\end{figure}

However, it is not straightforward to apply this interpretation to our stochastic model.
Note that $U$ does not have explicit time dependence in Sekimoto's definition. 
However, in our case, the energy of a Brownian particle in the backward process $U^*$ which is defined by 
${\bf u}^* ({\bf r},a,t) = -\nabla U^* ({\bf r},a,t)$,  
has an explicit time dependence because 
of the time dependence of $\rho$ (See Eq. (\ref{cc})).
To apply Sekimoto's procedure to our stochastic model, 
we introduce ``averaged" thermodynamic quantities and discuss the thermodynamic relation for these quantities.

Note that the force which acts on a Brownian particle at $t$ is given by 
${\bf u}({\bf X}(t),t) dt$ for the forward process and $-{\bf u}^* ({\bf X}(t),t) dt$ for the backward process, 
as is shown in Fig. \ref{fig1} ($dt > 0$).
Then we can introduce an averaged force which acts on a Brownian particle at $t$ as 
\begin{equation}
\bar{\bf u}({\bf X}(t),a,t) dt= {\bf b}({\bf X}(t),a,t) dt= \frac{{\bf u}({\bf X}(t),a,t) -{\bf u}^*({\bf X}(t),a,t)}{2} dt.
\end{equation}
Correspondingly, the averaged energy is defined as 
\begin{equation}
\bar{\bf u}({\bf X}(t),a,t) = - \nabla \bar{U} ({\bf X}(t),a,t),~~~~
\bar{U} ({\bf X}(t),a,t)= \frac{U({\bf X}(t),a,t) -U^*({\bf X}(t),a,t)}{2} .
\end{equation}

Following the discussion by Sekimoto, we implement the stochastic differentiation of $\bar{U}$ 
by using the Ito formula and obtain the thermodynamic relation for 
this averaged thermodynamic quantities as 
\begin{eqnarray}
d \bar{U} ({\bf X}(t),a,t) 
= \delta \bar{Q} ({\bf X}(t),a,t) + \delta \bar{\cal W} 
({\bf X}(t),a,t) + \delta Q_{J}({\bf X}(t),a,t), \label{ext_ther1}
\end{eqnarray}
where
\begin{eqnarray}
\delta \bar{Q} ({\bf X}(t),a,t) 
&=& \frac{\nabla U ({\bf X}(t),a,t) - \nabla U^* ({\bf X}(t),a,t) }{2}\circ d{\bf X}(t),  \\
\delta \bar{\cal W} ({\bf X}(t),a,t) 
&=& \frac{\partial_a U({\bf X}(t),a,t) - \partial_a U^* ({\bf X}(t),a,t)}{2}\circ da, \\
\delta  Q_{J}({\bf X}(t),a,t)  
&=& -\beta^{-1} \partial_t \ln \rho ({\bf X}(t),a,t) dt \nonumber \\
&=& -\beta^{-1}e^{\beta \bar{U} ({\bf X}(t),a,t)}\nabla \cdot {\bf J}({\bf X}(t),a,t) dt. \label{dqj_def}
\end{eqnarray}
Note that ${\bf X}(t)$ is a stochastic variable, 
we have to use the Ito formula instead of the Taylor expansion \cite{gardiner}.
In the derivation of the third equation, we used the macroscopic current defined from the Fokker-Planck equation as, 
\begin{equation}
\partial_t \rho({\bf r},a,t) = \nabla \cdot {\bf J}({\bf r},a,t).
\end{equation}
One can see that the first two terms on the r.h.s. of Eq. (\ref{ext_ther1}) already appear 
in Sekimoto's thermodynamics relation, and 
$\delta \bar{Q}$ and $\delta \bar{\cal W}$ are interpreted as the averaged heat and the averaged work, 
respectively. 
However, the last term does not appear in Eq. (\ref{thermo_seki}).

In order to clarify the meaning of the new term $\delta Q_{J}$, let us consider the one-dimensional free Brownian motion.
Then  $\delta Q_{J}$ is calculated as
\begin{equation}
\delta Q_J (X(t),t)
= -\beta^{-1} \frac{\partial_x J(X(t),t)}{\rho(X(t),t)} dt 
= \left( \frac{\beta^{-1}}{2t} - \frac{X^2(t)}{4t^2} \right) dt. \label{eqn:wj_brown}
\end{equation}
It is known that 
the expectation value of the position of a Brownian particle is given by $\langle X^2(t) \rangle = 2\beta^{-1}t$, and 
one can easily see that $\delta Q_J$ vanishes in this case.
Thus this term should be interpreted as another heat which a Brownian particle receives from the environment; 
when $X(t)$ moves slower than the average, 
$\delta Q_J$ is positive and a Brownian particle recieves heat, and vice versa.
Apparently from Eq. (\ref{dqj_def}), this heat appears only when $\rho$ has an explicit time dependence, that is, 
there is a non-vanishing macroscopic flow ${\bf J}$.

Note that, for the free Brownian motion, the energy of the forward process $U$ does not exist, but 
the averaged energy $\bar{U}$ is still finite.
By using Eqs. (\ref{cc}) and (\ref{dm_brown}), the averaged energy is calculated as 
\begin{equation}
\bar{U} (X(t),t) = \frac{X^2(t)}{4t} + \frac{\beta^{-1}}{2}\ln (4\pi \beta^{-1} t).
\end{equation}
Then the extended thermodynamic relation (\ref{ext_ther1}) for the free Brownian motion is explicitly given by 
\begin{equation}
d\left( \frac{X^2(t)}{4t} + \frac{\beta^{-1}}{2}\ln (4\pi \beta^{-1} t) \right) 
= \left( \frac{\beta^{-1}}{2t} - \frac{X^2(t)}{4t^2} \right) dt + \frac{X(t)\circ dX(t)}{2t}.
\end{equation}
The last term on the r.h.s. comes from $\delta \bar{Q}$,  
and there is no contribution from $\delta \bar{\cal W}$ in this case.

It should be noted that $\delta Q_J$ is re-expressed as,  
\begin{eqnarray}
\delta Q_{J}({\bf X}(t),a,t)
&=& -\beta \nabla (e^{\beta \bar{U} ({\bf X}(t),a,t)}{\bf J}({\bf X}(t),a,t) ) dt \nonumber \\
&& - {\bf J}({\bf X}(t),a,t) \cdot {\bf b}({\bf X}(t),a,t) e^{\beta \bar{U} ({\bf X}(t),a,t)} dt. 
\label{dwj2}
\end{eqnarray}
The first term on the r.h.s. is a total differentiation and may vanish for the total amount of  
the thermodynamic quantities. 
Note that ${\bf b}$ defined in Eq. (\ref{def_b}) is the averaged force on a Brownian particle 
and the second term is expressed as the product of the macroscopic current ${\bf J}$ and the microscopic force ${\bf b}$.
Remember that in the irreversible thermodynamics, the entropy production $\sigma$ is assumed to be given by 
the product of an irreversible current ${\bf J}_{irr}$ and the corresponding thermodynamic force ${\bf F}$, 
\begin{equation}
\sigma = {\bf J}_{irr} \cdot {\bf F}.
\end{equation}
One can see that the form of the second term of Eq. (\ref{dwj2}) is very similar to this structure.
In a quasi-static process, we may be possible to identify 
\begin{eqnarray}
\delta \bar{Q} &=& T dS, \\
\delta \bar{\cal W} &=& -PdV.
\end{eqnarray}
Then the extended thermodynamic relation is rewritten as
\begin{equation}
d \bar{U} = \delta Q_{J} + T dS - PdV.
\end{equation}
This is very similar to the generalized thermodynamic relation discussed in 
the extended irreversible thermodynamics \cite{jou}.

\section{fluctuation theorem in stochastic processes}

In this section, we will discuss the various fluctuation theorems in our stochastic model.
Note that the fluctuation theorems based on the stochastic dynamics have already been discussed 
in several works \cite{seifert,hatano}.
Differently from these works, our fluctuation theorems are expressed in terms of the averaged thermodynamic quantities 
introduced in the previous section.

Now we consider a set of a stochastic trajectory $\{ {\bf X}(t) \}$ 
which is described by Eq. (\ref{langevin1}).
Then, by calculating explicitly the total differentiation 
of $\bar{U}$ along this trajectory using the Ito formula \cite{gardiner}, 
we find the following relation,
\begin{equation}
d \bar{U} ({\bf X}(t),a, t) = (dU({\bf X}(t),a, t)- dU^*({\bf X}(t),a, t))/2 = -\beta^{-1} d\ln \rho ({\bf X}(t),a, t) 
,\label{eqn:u_dlnr}
\end{equation}
where
\begin{eqnarray}
d \ln \rho ({\bf X}(t),a,t) 
&=&
[\partial_t  + {\bf u}({\bf X}(t),a,t) \nabla + \beta^{-1} \nabla^2  ] \ln \rho ({\bf X}(t),a,t) dt 
+ \partial_a \ln \rho ({\bf X}(t),a,t) \circ da \nonumber \\
&& + \sqrt{2\beta^{-1}}\nabla \ln \rho ({\bf X}(t),a,t) \cdot d{\bf W}(t). \label{ours1}
\end{eqnarray}
Here the product in the inner product of the last term
 obeys the Ito definition \cite{gardiner}.

Thus, for the evolution from $t_0$ to $t_f$, we have
\begin{equation} 
\frac{\rho ({\bf X}(t_f),a_f, t_f)}{\rho ({\bf X}(t_0),a_0, t_0)} 
= e^{-\beta (\bar{U} ({\bf X}(t_f),a_f, t_f) - \bar{U} ({\bf X}(t_0),a_0, t_0) )}. \label{exact1} 
\end{equation}

\subsection{Quasi-static change between two different stationary states}

For the quasi-static process where we assume $\partial_t \rho ({\bf r},a,t) \approx 0$, 
the extended thermodynamic relation is approximately given by  
\begin{eqnarray}
d \bar{U} ({\bf X}(t),a,t) \approx \delta \bar{Q} ({\bf X}(t),a,t)+ \delta \bar{\cal W} ({\bf X}(t),a,t) .
\end{eqnarray}
Thus, from Eq. (\ref{exact1}), the two stationary state are connected by the following relation,
\begin{eqnarray}
&& \frac{\rho_{sta} ({\bf X}(t_f),a_f)}{\rho_{sta} ({\bf X}(t_0),a_0)} = 
e^{-\beta \sum_{i=0}^{f-1} ( \delta \bar{Q} ({\bf X}(t),a,t)+\delta \bar{\cal W} ({\bf X}(t),a,t) )}. 
\end{eqnarray}
Clearly, there is no contribution from the macroscopic flow, $\delta Q_J$, in this case.

\subsection{Seifert relation}

We consider a stochastic trajectory $\Gamma = \{ {\bf X}_0, {\bf X}_1, \cdots, {\bf X}_f  \}$ 
which is a solution of the stochastic difference equations (\ref{langevin1}), 
and introduce the corresponding time-reversed process denoted by $\Gamma^{\dagger}$.
Then we consider the conditional probability which is defined by
\begin{equation}
P({\bf X}_j,a_j,t_j|{\bf X}_i,a_i,t_i) = \frac{P ( ({\bf X}_j,a_j,t_j) \cap ({\bf X}_i,a_i,t_i))}{P({\bf X}_i,a_i,t_i)} 
= \frac{\rho({\bf X}_j,a_j,t_j)\rho({\bf X}_i,a_i,t_i)}{\rho({\bf X}_i,a_i,t_i)} = \rho({\bf X}_j,a_j,t_j),
\end{equation}
where  $\rho$ is the solution of the Fokker-Planck equation (\ref{fp1}).
Here we used that the process is Markovian.
From Eq. (\ref{exact1}), the r.h.s. of the above equation is given by 
\begin{equation}
\rho ({\bf X}_j,a_j,t_j) 
= \rho({\bf X}_i,a_i,t_i) e^{-\beta ( \bar{U} ({\bf X}_j,a_j,t_j) - \bar{U} ({\bf X}_i,a_i,t_i) ) } 
= e^{-\beta  \bar{U} ({\bf X}_j,a_j,t_j) } .
\end{equation}
The transition probability from an initial ${\bf X}_0$ to a final ${\bf X}_f$ along the stochastic trajectory $\Gamma$ 
is, finally, given by 
\begin{eqnarray}
\lefteqn{P(\Gamma ({\bf X}_0,a_0,t_0;{\bf X}_f,a_f,t_f)) } && \nonumber \\
&\equiv& P({\bf X}_f,a_f,t_f|{\bf X}_{f-1},a_{f-1},t_{f-1})P({\bf X}_{f-1},a_{f-1},t_{f-1}|{\bf X}_{f-2},a_{f-2},t_{f-2})
\cdots P({\bf X}_1,a_1,t_1|{\bf X}_0,a_0,t_0) \nonumber \\
&=& e^{-\beta [\sum_{i=0}^f \bar{U} ({\bf X}_i,a_i,t_i) - \bar{U} ({\bf X}_0,a_0,t_0) ] }.
\end{eqnarray}
Similarly, the conditional probability for the inverse process is 
\begin{eqnarray}
\lefteqn{P (\Gamma^{\dagger} ({\bf X}_0,a_0,t_0;{\bf X}_f,a_f,t_f)) } && \nonumber \\
&\equiv & P({\bf X}_0,a_0,t_0|{\bf X}_{1},a_1,t_{1})P({\bf X}_{1},a_1,t_{1}|{\bf X}_{2},a_2,t_{2})
\cdots P({\bf X}_{f-1},a_{f-1},t_{f-1}|{\bf X}_f,a_f,t_f)  \nonumber \\
&=&  e^{ -\beta [ \sum_{i=0}^f \bar{U} ({\bf X}_i,a_i,t_i) - \bar{U} ({\bf X}_f,a_f,t_f) ] }.
\end{eqnarray}

Following the discussion by Seifert \cite{seifert},
we introduce $R(\Gamma ({\bf X}_0,a_0,t_0;{\bf X}_f,a_f,t_f))$ as follows,
\begin{eqnarray}
\lefteqn{R(\Gamma ({\bf X}_0,a_0,t_0;{\bf X}_f,a_f,t_f)) } && \nonumber \\
&=& 
\ln \frac{P(\Gamma ({\bf X}_0a_0,,t_0;{\bf X}_f,a_f,t_f))\rho ({\bf X}_0,a_0,t_0)}
{P (\Gamma^{\dagger} ({\bf X}_0,a_0,t_0;{\bf X}_f,a_f,t_f))\rho ({\bf X}_f,a_f,t_f)} \nonumber \\
&=&
\ln \frac{\rho ({\bf X}_0,a_0,t_0)}{\rho ({\bf X}_f,a_f,t_f)}
-\beta (\bar{U}({\bf X}_f, a_f, t_f) - \bar{U}({\bf X}_0, a_0, t_0) ) \nonumber \\
&=&
\ln \frac{\rho ({\bf X}_0,a_0,t_0)}{\rho ({\bf X}_f,a_f,t_f)} -\beta \sum_{i=0}^{f-1} (\delta Q_J ({\bf X}_i,a_i,t_i)
+\delta \bar{Q} ({\bf X}_i,a_i,t_i) + \delta \bar{\cal W} ({\bf X}_i,a_i,t_i)). \label{r_def} 
\end{eqnarray}

By using this, we can derive the exact relation,
\begin{eqnarray}
\lefteqn{\langle e^{-R(\Gamma ({\bf X}_0,a_0,t_0;{\bf X}_f,a_f,t_f))} \rangle } && \nonumber \\
&\equiv&  \int d^3 {\bf X}_0d^3 {\bf X}_f \langle P(\Gamma ({\bf X}_0,a_0,t_0;{\bf X}_f,a_f,t_f))\rho ({\bf X}_0,a_0,t_0) 
e^{-R(\Gamma ({\bf X}_0,a_0,t_0;{\bf X}_f,a_f,t_f))} \rangle_W \nonumber \\
&=& 1, \label{oursei}
\end{eqnarray} 
were $\langle~\rangle_W$ means the average over possible trajectories 
fixing ${\bf X}_0$ at $t_0$ and ${\bf X}_f$ at $t_f$.
Here we used the following relations,
\begin{eqnarray}
\rho ({\bf X}_0,a_0,t_0) &=& \int d^3 {\bf X}_f \langle P(\Gamma^{\dagger} ({\bf X}_0,a_0,t_0;{\bf X}_f,a_f,t_f))\rho ({\bf X}_f,a_f,t_f) \rangle_W, \\
 \rho ({\bf X}_f,a_f,t_f) &=& \int d^3 {\bf X}_0 \langle P(\Gamma ({\bf X}_0,a_0,t_0;{\bf X}_f,a_f,t_f))\rho ({\bf X}_0,a_0,t_0) \rangle_W .
\end{eqnarray}
This is the main result, leading to various fluctuation theorems.

The expression (\ref{oursei}) corresponds to the relation discussed by Seifert \cite{seifert}, 
but the expression of $R$ is different.
Seifert gives the following representation for $R$,  
\begin{equation}
R_{\mathrm{seifert}} 
= \ln \frac{\rho ({\bf X}_0,a_0,t_0)}{\rho ({\bf X}_f,a_f,t_f)} + \beta \sum_{i=0}^{f-1} {\bf u}_i \circ d{\bf X}_i 
=  \ln \frac{\rho ({\bf X}_0,a_0,t_0)}{\rho ({\bf X}_f,a_f,t_f)} -\beta \sum_{i=0}^{f-1} \delta Q({\bf X}_i,a_i,t_i).\label{r_sei}
\end{equation}
Thus our exact relation (\ref{oursei}) is not equivalent to the Seifert relation.

The validity of the Seifert relation was numerically checked \cite{test}. 
However, our result is also consistent with this numerical result.
In \cite{test}, 
the stationary state with fixed $a$ is considered.
In this limited situation, our averaged quantities are reduced to 
\begin{eqnarray}
\delta Q_J = \delta \bar{\cal W} = 0,~~~~\delta \bar{Q} =  \delta Q.
\end{eqnarray}
Then our $R$ in (\ref{r_def}) becomes equivalent to $R_{\mathrm{seifert}}$.
In this sense, our result is still consistent with \cite{test}.

\subsection{Jarzynski relation}

It is easy to derive the relation corresponds to the Jarzynski relation \cite{jar} from our relation (\ref{oursei}).
We consider that the evolution of a Brownian particle from one stationary state $\rho_{sta}({\bf r},a_0)$,  
to the other one $\rho_{sta}({\bf r},a_f)$.
The stationary state is given by 
\begin{equation}
\rho_{sta} ({\bf r},a) = \frac{1}{Z_{sta}(a)} e^{-\beta U({\bf r},a)},
\end{equation}
where $Z_{sta}(a) = \int d^3{\bf r} e^{-\beta U({\bf r},a)}$.
Substituting it into Eq. (\ref{oursei}), we obtain the Jarzynski relation of our stochastic model,
\begin{equation}
\langle e^{-\beta(U({\bf X}_f,a_f) - U({\bf X}_0,a_0) -\beta \sum_{i=0}^{f-1} (\delta Q_J ({\bf X}_i,a_i,t_i)
+\delta \bar{Q} ({\bf X}_i,a_i,t_i) + \delta \bar{\cal W} ({\bf X}_i,a_i,t_i)) )} \rangle
= e^{-\beta \Delta F},
\end{equation}
where 
\begin{equation}
\Delta F = -\beta^{-1} \ln \frac{Z_{sta}(a_f)}{Z_{sta}(a_0)}.
\end{equation}

\subsection{Komatsu-Nakagawa non-equilibrium steady state}

Recently, the general expression of non-equilibrium steady state was derived in \cite{kn,knst}.
The similar expression can be obtained in our stochastic model.

From Eq. (\ref{r_def}), we have
\begin{eqnarray}
\lefteqn{P(\Gamma^{\dagger} ({\bf X}_0,a_0,t_0;{\bf X}_f,a_f,t_f))\rho ({\bf X}_f,a_f,t_f) } && \nonumber \\
&=&   P(\Gamma ({\bf X}_0,a_0,t_0;{\bf X}_f,a_f,t_f))\rho ({\bf X}_0,a_0,t_0) e^{-R(\Gamma ({\bf X}_0,a_0,t_0;{\bf X}_f,a_f,t_f))} .
\end{eqnarray}
Similarly, for the inverse process, we obtain 
\begin{eqnarray}
\lefteqn{P(\Gamma ({\bf X}_0,a_0,t_0;{\bf X}_f,a_f,t_f))\rho ({\bf X}_0,a_0,t_0) } && \nonumber \\
&=&   P(\Gamma^{\dagger} ({\bf X}_0,a_0,t_0;{\bf X}_f,a_f,t_f))\rho ({\bf X}_f,a_f,t_f) e^{-R(\Gamma^{\dagger} ({\bf X}_0,a_0,t_0;{\bf X}_f,a_f,t_f))} .
\end{eqnarray}
By combining these equations, we have
\begin{eqnarray}
\lefteqn{P(\Gamma ({\bf X}_0,a_0,t_0;{\bf X}_f,a_f,t_f))\rho ({\bf X}_0,a_0,t_0) e^{-R(\Gamma ({\bf X}_0,a_0,t_0;{\bf X}_f,a_f,t_f))/2} } && \nonumber \\
&&= P(\Gamma^{\dagger} ({\bf X}_0,a_0,t_0;{\bf X}_f,a_f,t_f))\rho ({\bf X}_f,a_f,t_f) e^{-R(\Gamma^{\dagger} ({\bf X}_0,a_0,t_0;{\bf X}_f,a_f,t_f))/2}. 
\end{eqnarray}
This is re-expressed as 
\begin{eqnarray}
\rho ({\bf X}_f,a_f,t_f) = \rho ({\bf X}_0,a_0,t_0)  
\frac{\langle P(\Gamma ({\bf X}_0,a_0,t_0;{\bf X}_f,a_f,t_f)) e^{-R(\Gamma ({\bf X}_0,a_0,t_0;{\bf X}_f,a_f,t_f))/2}\rangle_W}
{\langle  P(\Gamma^{\dagger} ({\bf X}_0,a_0,t_0;{\bf X}_f,a_f,t_f)) e^{-R(\Gamma^{\dagger} ({\bf X}_0,a_0,t_0;{\bf X}_f,a_f,t_f))/2} \rangle_W}
\end{eqnarray}
Note that ${\bf X}_f$ and ${\bf X}_0$ are fixed parameters and independent of the average $\langle ~ \rangle_W$.
This is essentially the same as the expression derived in \cite{kn} (Eq. (10)), except for the definition of $R$. 
The quantity $\hat{\Theta}_I$ in \cite{kn}, which corresponds to $R$ in our case, coincides with 
Eq. (\ref{r_def}) when $\delta Q_J$ vanishes.
That is, the Komatsu-Nakagawa non-equilibrium steady state is realized in our stochastic model only 
when there is no macroscopic flow.

\section{Concluding remarks}

In this paper, we considered the stochastic model incorporating the forward process and the backward process in time.
Following the strategy of Sekimoto \cite{seki}, we constructed the thermodynamic relation for 
the ``averaged" quantities of the forward and backward processes. 
Our thermodynamic relation is extended so that a term $\delta Q_J$ associated with a macroscopic flow appears.
We further discussed the fluctuation theorems in our stochastic model and derived the new expressions 
with respect to the averaged quantities.

The new term $\delta Q_J$ can be expressed as the product of a macroscopic current and a force.
This is the form expected from the extended irreversible thermodynamics. 
This term may have a relation to the excess heat discussed in \cite{oono}

The consideration of the forward and backward stochastic equations is, in fact, well 
known in Nelson's formulation of quantum mechanics, which is one of the hidden variable theories \cite{nelson}.
As a matter of fact, we can introduce a kind of a wave function even in the classical Brownian motion and 
it is possible to interpret the fluctuation theorems in terms of the phase of the wave function.
This will be reported in another work.

In this work, we discussed the stochastic equation of motion without inertial terms under a Gaussian white noise.
It is also interesting to extend our discussion to the equations with inertial terms, memory effects and colored noise.
In such a situation, it may be possible to investigate the dynamics of the pre-equilibrium state 
as is discussed in \cite{kk}.
Moreover, if our discussion is applicable to relativistic Brownian motion \cite{hanggi}, 
we can investigate the thermodynamic relation in relativistic systems.

T. Koide acknowledges useful comments by G. S. Denicol.
This work was (financially) supported by the Helmholtz International
Center for FAIR within the
framework of the LOEWE program (Landesoffensive zur Entwicklung
Wissenschaftlich-
Okonomischer Exzellenz) launched by the State of Hesse.

\end{document}